\documentclass[%
reprint,
 amsmath,amssymb,
 aps,
floatfix
]{revtex4-1}
\usepackage{color}
\usepackage{graphicx}
\usepackage{dcolumn}
\usepackage{bm}
\usepackage{ulem}

\begin{document}

\title{Electron Correlation in the Lanthanides: $4f^2$ spectrum of Ce$^{2+}$}

\author{Charlotte Froese Fischer}
 \email{cff@cs.ubc.ca}
\affiliation{Department of Computer Science, University of British Columbia, 2366 Main Mall,
Vancouver, BC V6T1Z4, Canada}
\author{Michel R. Godefroid}
\affiliation{Chimie Quantique et Photophysique, CP160/09 \\ Universit\'e libre de Bruxelles, 1050 Brussels, Belgium}

\date{\today}

\begin{abstract}
Atoms and ions of Lanthanides have multiple opens shells along with an open $4f^k$ subshell. This paper studies the effect of electron correlation in such systems and how wave functions can be determined
for the accurate prediction of atomic properties in the case of Ce$^{2+}$ where $k=2$, using the
multireference single- and double-excitation method. An efficient higher-order  method is recommended for more reliable results.
\end{abstract}

\maketitle


\section{Introduction}

Lanthanides were detected recently in the electromagnetic counterpart to a gravitational wave source from a binary neutron star merger (GW170817)~\cite{Cowetal:2017a}.
Knowledge of their atomic structure is essential for estimating the ejecta opacity and  understanding the r-nuclear process at the origin of their synthesis~\cite{Piaetal:2017a,Kasetal:2017a}.
Lanthanides and related Actinides are also the elements of the periodic table that pave the way to 
the transfermium elements ($Z \geq 100$) that do not occur naturally on Earth and are produced at large accelerator facilities,  
for which the atomic structure is almost unknown~\cite{Laaetal:2016a}, and to super-heavy elements ($Z \geq 104$) that are good candidates for the island of stability of nuclear astrophysics interest~\cite{Dzuetal:2017b}.  \\

To estimate the r-process opacities that are dominated by bound-bound transitions,  the radiative transition rates  have to be calculated for tens of millions of lines in lanthanide ions, using atomic structure models that   determine the approximate ion energy level structure and the wavelengths and oscillator strengths of all permitted radiative transitions \cite{Kasetal:2013a}. Although these models do not provide exact results, the hope is that they capture the statistical distribution of levels and lines, to derive reliable estimates of the pseudo-continuum opacity~\cite{Fonetal:2017b}. Benchmark calculations for a few elements have been performed \cite{Tanetal:2018a} to confirm that 
 the opacities from bound-bound transitions of open $f$-shell elements are higher than those of the other elements over a wide wavelength range. 
The present work does not enter in this category of  calculations. It mainly focuses on the search of the relevant correlation configurations entering in the description of atomic energy levels of complex atomic systems and to the development of {\it ab initio} computational strategies allowing their efficient inclusion. The ultimate goal is to improve the reliability of theoretical atomic energy levels, excitation energies and wave function compositions, in line of other recent works~\cite{FroGai:2018a,Gedetal:2018a}. \\

Parametric studies can be performed to unravel the complex spectra of Lanthanides (and Actinides) (see for instance~\cite{WyaPal:98a} for Ce$^{2+}$)  but 
needed are the observed atomic line frequencies and intensities, which are precisely the targets of {\it ab initio} approaches. 
The effect of correlation in atoms and ions of Lanthanides and Actinides is not well understood.  
Safronova~{\it et al.}~\cite{Safetal:2015a} summarize the situation well-- ''though tremendous progress has been made, calculations for the Lanthanides with the open $4f$-shell  remain a challenge.'' 
In their paper, they report results from applying two hybrid approaches to the elements La, La$^+$, Ce, Ce$^+$, Ce$^{2+}$, and Ce$^{3+}$. In their studies, not all levels of a configuration are included. In particular, in Ce$^{2+}$ ($Z=58$) only five levels were reported, namely $^3H_{4,5,6},\; ^1G_4,\;^1D_2$ instead of the thirteen levels arising from a single open subshell $f^2$ configuration~\cite{Jud:98a}. Their method is based on the use of an effective Hamiltonian for including correlation within the closed subshells and configuration interaction (CI) for electrons in open subshells (referred to as valence electrons)  and perturbation theory methods of various orders. 

The present paper discusses similar strategies based on variational methods for determining wave functions that can be used to predict atomic properties and not only energies, methods that have been implemented in the 
 General Relativistic Atomic Structure Package computer codes
 ({\sc Grasp}2K~\cite{Jonetal:2013a} and {\sc Grasp}2018~\cite{GRASP2018}).
 
  What makes the calculations challenging is the rapid explosion in the number of basis states associated with configurations with multiply occupied subshells with large angular momenta and the need for higher-order corrections.
  The configuration [Kr]$4d^84f^45s^25p^45d^2$ of Ce$^{2+}$ has associated with it 1,608,502 basis states, for 
  $0 \le J \le 6$.
  In addition, strong interactions require treatments for higher-order effects and standard procedures rapidly produce expansions of 10~Million basis states or more. Once wave functions have been determined other properties can be computed.
 
\section{Underlying theory} 

In the multiconfiguration Dirac-Hartree-Fock (MC-DHF) method~\cite{Froetal:2016a}, the wave function $\Psi({\mathit \gamma} \pi JM_J)$ for a
state labeled ${\mathit \gamma} \pi JM_J$, where  $J$ and $M_J$ are the angular quantum numbers and $\pi$ the
parity, is expanded in antisymmetrized 
 configuration state functions (CSFs) 
\begin{equation}
\label{ASF}
\Psi({\mathit \gamma} \pi JM_J)  = \sum_{j=1}^{N} c_{j} \Phi(\gamma_{j} \pi JM_J).
\end{equation}
The labels $\{\gamma_j\}$ denote other appropriate information about the CSFs,
such as orbital occupancy and the subshell coupling tree.
The  CSFs are built from products of one-electron orbitals, having the general form
\begin{align}
\label{g2}
\psi_{n\kappa, m}(\mathbf{r}) = \frac{1}{r}
\begin{pmatrix} P_{n\kappa}(r) \chi_{\kappa, m}(\theta,\varphi) \\
i Q_{n\kappa}(r) \chi_{-\kappa, m}(\theta,\varphi)
\end{pmatrix},
\end{align}
where $\chi_{\pm \kappa, m}(\theta,\varphi)$ are 2-component spin-angular functions. 
The radial functions $\{P_{n\kappa}(r), Q_{n\kappa}(r)\}$ are  represented numerically on a grid.

Radial functions are solutions of systems of differential equations that define a stationary state of an energy functional for one or more wavefunction expansions. It is possible to derive the MCDHF equations from the usual variational procedure by varying both the large and small component so that
 \begin{eqnarray}
 \label{eq:DHF-operator}
 \hspace{-2cm}
 w_a
 \left[ \begin{array}{c c}
   V(a;r)  & 
  -c\left[ \frac{d}{dr} - \frac{\kappa_a}{r} \right] \\
      \ \\
      c\left[ \frac{d}{dr} + \frac{\kappa_a}{r} \right] & V(a;r)  -2c^2
      \end{array}\right]
    \left[\begin{array}{c}
       P_a(r)\\
       Q_a(r)
      \end{array}\right] \nonumber \\
       = \sum_{b} \epsilon_{ab} \;  \delta_{\kappa_a \kappa_b}
 \left[\begin{array}{c}
       P_b(r)\\
       Q_b(r)
       \end{array}\right],
\end{eqnarray}
 where $V(a;r) = V_{nuc}(r) + Y(a;r) + \bar{X}(a;r)$ is a potential consisting of nuclear, direct, and exchange contributions  arising from both diagonal and off-diagonal matrix elements, $\langle \Phi_\alpha \vert {\cal H}_{DC}  \vert \Phi_\beta \rangle$, of the Dirac-Coulomb (DC) Hamiltonian~\cite{Froetal:2016a}. In each $\kappa$-space, Lagrange related energy parameters $\epsilon_{ab} = \epsilon^{\kappa}_{n_a n_b}$ are introduced to impose orthonormality constraints in the variational process. 
In spectrum calculations, where only energy differences relative to the ground state are important, wave functions for a number of targeted states are determined simultaneously in the extended optimal level (EOL) scheme. This assures that different eigenstates of the symmetry are orthonormal even though the solutions are approximate.

Given initial estimates of the radial functions, the energies $E$ and expansion coefficients ${\bf c} = (c_1,\ldots,c_N)^t$ for the targeted states are obtained as solutions to the configuration interaction (CI) problem,
\begin{equation}
{\bf H}{\bf c} = E {\bf c},
\end{equation}
where  ${\bf H}$ is the  CI matrix of dimension $N \times N$ with elements
\begin{equation}
H_{ij} = \langle \Phi(\gamma_{i} \pi JM_J)|H| \Phi(\gamma_{j} \pi JM_J) \rangle.
\end{equation}

In {\sc Grasp}, expansions in terms of CSFs are obtained through single- and double-excitations (SD) from a multireference~(MR) set of CSFs that contain the important contributions to the wave function composition. In systematic calculations the excitations are to  orbital sets of increasing size that include both unfilled and virtual orbitals. Calculations often are classified by their maximum principal quantum number so that an $n=5$ calculation has associate with it excitations to all orbitals up to $5g$. When the orbital set is increased in size, only the new orbitals need be determined. Expansions may grow rapidly in size, so partitioning CSFs   and omitting interactions between  new CSFs can drastically reduce the computation in the self-consistent process.  

A {\sc Grasp} calculation consists of three phases -- i) generating the expansions, ii) building the orbital basis using variational theory for the Dirac-Coulomb Hamiltonian, and iii) performing a relativistic configuration interaction calculation that includes the transverse photon and QED corrections. This process is described in detail in the manual for {\sc GRASP2018}~\cite{GRASP2018_manual_git}. 

\section{Large expansions}

When expansions become exceedingly large which is the case when millions of small effects (small expansion coefficients) are present, it is useful to partition the set of CSFs according to some criterion to produce a zero-order set and a first-order correction, respectively~\cite{Gusetal:2017a}. Suppose the expansion coefficients were vectors $c^{(0)}$ and $c^{(1)}$, respectively. 
This partitioning also  divides the interaction matrix $H$ into blocks so that the eigenvalue problem becomes
\begin{equation}
\label{eq:partitionCI}
\left( \begin{array}{cc} H^{(00)} & H^{(01)} \\ H^{(10)} & H^{(11)} \end{array} \right) 
\left( \begin{array}{c} c^{(0)}  \\ c^{(1)}  \end{array} \right)
=  E
\left( \begin{array}{c} c^{(0)} \\ c^{(1)}  \end{array} \right), 
\end{equation}\\
where $H^{(00)}$ is the interaction matrix between zero-order components, $H^{(11)}$ for interactions between first-order components of the wave function, 
and $H^{(01)}=H^{(10)\dagger}$ 
represents the interactions between CSFs of the two blocks.
This equation can be rewritten as a pair of linear equations, namely
\begin{equation}
\begin{array}{cccl}
( H^{(00)}-EI) c^{(0)} &+&  H^{(01)} c^{(1)} & = 0, \\
 H^{(10)} c^{(0)} &+&   (  H^{(11)} - EI ) c^{(1)} &  =0.  \end{array}
\end{equation}
Solving for $c^{(1)}$ in the second equation and substituting into the first, we get an eigenvalue 
problem for $c^{(0)}$, 
\begin{equation}
\label{eq:deflate}
\hspace*{-0.75cm}
\small{
\left( H^{(00)} - H^{(01)}\left(H^{(11)}-EI\right)^{-1}H^{(10)} - EI \right) c^{(0)} =0.
}
\end{equation}\\
This deflates the matrix in that it  reduces the eigenvalue problem  for a matrix of size $N\times N$ (several million) to an eigenvalue problem of size $m \times m$ (several tens of thousands), where $m$ is the expansion size of $c^{(0)}$. Of course, once  $E$ and $c^{(0)}$ have been determined, the other components can be generated from the expression 
\begin{equation}
\label{eq:c1}
c^{(1)} = -(H^{(11)}-EI)^{-1}H^{(10)} c^{(0)},
\end{equation}
and a full wave function is defined. 
Note that the eigenvalue problem is now non-linear in the eigenvalue that can be solved by an iterative process.  When $H^{(11)}-EI$ is replaced by the diagonal matrix such as $H^{(11)}_{ii}-E^0I$, Eq.~\ref{eq:deflate} is again a linear eigenvalue problem. 

In the CI+MBPT approach~\cite{Kozetal:2015a,Dzuetal:2017a}, when $c^{(1)}$ is associated with correlation in the core, and $c^{(0)}$ with valence correlation, the matrix of Eq.~\ref{eq:deflate} represents the matrix from an effective Hamiltonian. Consequently, interactions between first-order core corrections to the wavefunction are not included. Thus, contributions to the wavefunction, need to be small. When other atomic properties are evaluated, it would be desirable for $c^{(1)}$ to be sufficiently small so that contributions from the relevant operator {\it between} small corrections can be omitted.

Partitioning the configuration interaction matrix so that the CSFs in $c^{(1)}$ space have small coefficients has been supported already in the {\sc Atsp} code~\cite{Froetal:2007a} but in
variational methods, omitting interactions between these CSF's comes at a cost.  The total energy associated with a wave function is an upper bound to the exact energy, but when off-diagonal matrix elements of $H^{(11)}$ are neglected, the  total energies often are too low.  In the present work, the final relativistic configuration interaction calculation always included the full matrix but used as many as 96 parallel processors for execution of the task.

Partitioning can also be introduced in the building of an orbital basis. Suppose the $n=5$ orbitals have already been determined and important contributors to the wave function composition have been identified.  These define $c^{(0)}$. Then the energy functional for the variational process could neglect interactions within the $ c^{(1)} $ space, greatly reducing the time for determine orbitals that satisfy orthogonality constraints.  Variational methods optimize the orbital basis. The effect on the calculation of neglecting some interactions is a slower rate of convergence of the systematic procedure and an extra layer
of orbitals may ultimately be needed.  
This process was used effectively in the study of Pr$^{3+}$~\cite{FroGai:2018a}. In the present study, this option was only used when expansions were large in which case the $c^{(0)}$ space was defined as the MR set, unless indicated otherwise.

\section{A two-electron system} 

\begin{figure}[t]
\begin{center}
\includegraphics[scale=0.25]{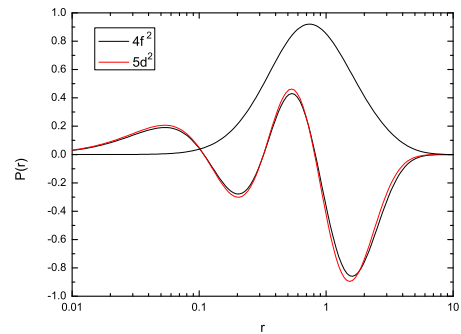}
\end{center}
\vspace{-4pt}
\caption{Figure showing the large components of $4f_+=4f_{7/2}$ (nodeless) and $5p_+=5p_{3/2}$ (oscillating) orbitals from the configurations $4f^25s^25p^6$ (black) and $5s^25p^65d^2$(red online, grey otherwise).}
\label{fig:5p-orb}
\end{figure}

A simple Dirac-Hartree-Fock calculation for the ground configuration [Xe]$4f^2$ of Ce$^{2+}$ shows that the $4f$ orbitals are not outer orbitals, but orbitals with mean radii  between those for
$\{4s,4p,4d\}$ and $\{5s,5p\}$ orbitals as shown in Table~\ref{tab:Ce-rad}. Results are given for two configurations, one with $4f^2$ and the other with $5d^2$. 
 \begin{table}
    \caption{Mean radii, $\langle r \rangle _{nl}$ (in $a_0$) of Ce$^{2+}$ orbitals for two configurations -- 1) $4f^25s^25p^6$ and 2) $5s^25p^65d^2$. Left column: $j=l-1/2~(\kappa > 0)$; right column: $j=l+1/2~(\kappa < 0)$. }
    \label{tab:Ce-rad}
    \begin{center}
\begin{tabular}{l r r c | c l r r}
\hline
  $nl$   &   $\langle r \rangle _{nl}$  &   $\langle r \rangle _{nl}$   & & & $nl$   &   $\langle r \rangle _{nl}$  &  $\langle r \rangle _{nl}$\\
  \hline
       &    1)  & 2) & \  &  & & 1) & 2) \\
\hline
       &        &       & & & $4s$   &  0.638 & 0.635 \\
  $4p_-$  &  0.659 & 0.657 & & & $4p_+$ &  0.684 & 0.679 \\
  $4d_-$  &  0.745 & 0.733 & & &  $4d_+$  &  0.757 & 0.742 \\
  $4f_-$  &  1.152 &       & & & $4f_+$  &  1.165 \\
          &        &       & & & $5s$    &  1.569 & 1.504 \\
  $5p_-$  &  1.752 & 1.659 & & & $5p_+$   &  1.830 & 1.727 \\
  $5d_-$  &        & 2.408 &~~~ &~~~ & $5d_+$   &        & 2.443 \\
\hline
\end{tabular}
\end{center}
\end{table}
Normally, for a given electron, the nucleus is screened by other electrons with a smaller mean radius. But Table~\ref{tab:Ce-rad} shows that when the $4f^2$ electrons are replaced by $5d^2$ electrons, the common orbital parameters hardly change.
Fig.~\ref{fig:5p-orb} shows how close to each other the large components of $5p$-orbitals of $4f^25s^25p^6$ (black, online and in text) and $5s^25p^65d^2$ (red online, grey in text) are. Also shown for the comparison is the nodeless $4f$ orbital. Because the $4f$ orbital amplitude is so small near the origin, it affects the potential for other electrons only at larger values of the radius.

By expanding the wave function for a two-electron system outside a core through SD excitations to an increasing set of orbitals, the $4f^2$ spectrum converges rapidly as shown in Table~\ref{tab:2-el}.  Because of the strong interaction between $4f^2$ and $5d^2$, radial functions were optimized (equally weighted) for levels of both configurations. For the converged results, the ground state energy ($E_g$) was $-8848.36$ $E_h$. 
For comparison, the observed energies from the Atomic Structure Database (ASD)~\cite{NIST_ASD} are provided as well as the best results reported by Safronova~{\it et al.}~\cite{Safetal:2015a}.
Note, however, that the $^1G_4$ level is not in the observed order.
This first analysis reveals the importance of the mixing of $4f^2 5s^2 5p^6$ with $5s^2 5p^6 5d^2$. \\
 \begin{table}[t]
\caption{$4f^2$ energy levels (in cm$^{-1}$) from a 2-electron calculation  compared with observed energy levels. The ground state energy is $-E_g = 8848.36$ $E_h$.}
\label{tab:2-el}
\begin{center}
\begin{tabular}{r r r r r r r}
\hline
   $LSJ$ &       $n=5$ &       $n=6$ &      $n=7$  &  $n=8$  &    ASD     &   CI all \\
         &             &             &             &         &  \cite{NIST_ASD} & \cite{Safetal:2015a} \\ 
\hline
    $^3H_4$ &         0 &         0 &         0  &      0  &       0.00  &    0 \\
    $^3H_5$ &      1246 &      1250 &      1249  &   1251  &    1528.32  &  1565 \\
    $^3H_6$ &      2571 &      2573 &      2567  &   2570  &    3127.10  &  3227 \\
    $^3F_2$ &      3870 &      3852 &      3808  &   3801  &    3762.75  &  \\
    $^3F_3$ &      4679 &      4663 &      4620  &   4614  &    4764.76  &  \\
    $^3F_4$ &      6399 &      6267 &      6206  &   6181  &    5006.06  &  \\
    $^1G_4$ &      4678 &      4510 &      4442  &   4403  &   7120.00  &  7650 \\
    $^1D_2$ &     13639 &     13316 &     13175  &  13103  &  12835.09   & 13786 \\
    $^3P_0$ &     17067 &     16944 &     16825  &  16807  &  16072.04  &  \\
    $^3P_1$ &     17485 &     17368 &     17253  &  17237  &  16523.66  &  \\
    $^3P_2$ &     18171 &     18037 &     17921  &  17903  &  17317.49  &  \\
    $^1I_6$ &     19668 &     19157 &     19104  &  19045  &  17420.60  &  \\
    $^1S_0$ &     32006 &     30967 &     30512  &  30362  &  ~~~~~32838.62  &  \\
\hline
\end{tabular}
\end{center}
\end{table}

\begin{table}[b]
    \caption{The wave function expansion for the largest basis states  of  the super-complex. Included are the CSFs, their expansion coefficient, and the excitation producing the CSF. }
    \label{tab:expand}
    \centering
    \begin{tabular}{l r l}
    \hline
    CSF & Coeff. & Excitation \\
    \hline
     $4d^{10}5s^25p^6$            & 0.9734 \\
     $4d^94f(^1\!P)5s^25p^55d$      & 0.0854  & $4d5p\;\rightarrow 4f5d$\\
     $4d^8(^3\!P)4f^2(^3\!P)5s^25p^6$ &$-$0.0715 & $4d^2\; ^3\!P \rightarrow 4f^2\; ^3\!P $\\
     $4d^8(^3\!F)4f^2(^3\!F)5s^25p^6$ &$-$0.0596 & $4d^2\; ^3\!F \rightarrow 4f^2\; ^3\!F $\\
     $4d^{10}5s^25p^4(^3\!P)5d^2(^3\!P)$&$-$0.0559 & $5p^2\;^3\!P\rightarrow 5d^2\; ^3\!P$\\
     $\ldots$ \\
     $4d^8(^3\! P)4f^2(^3\!P)5s^25p^4(^3\!P)5d^2(^3\!P)$\;&0.0059 &
     $4d^2\; ^3\!P \rightarrow 4f^2\; ^3\!P $\\
       & & $5p^2\;^3\!P\rightarrow 5d^2\; ^3\!P$\\
    \hline
    \end{tabular}
\end{table}

\section{Some properties of correlation}

The Lanthanides all have two incomplete shells, namely the $n=4$ shell that is missing $4f$ electrons, and the $n=5$ shell missing $5d, 5f, 5g$ electrons. Each of these shells have a complex of configurations that may interact
strongly through near degeneracy~\cite{LayBah:62a}. Let us consider Ce$^{4+}$ where all subshells are filled.  In this case the complexes are denoted as $\{ 4s+4p+4d \}^{18}=4^{18}$ and $\{ 5s + 5p \}^8=5^8$, respectively, where the exponent denotes the number of electrons in a given shell. These two complexes can be merged into a {\it super-complex}~$4^{18}5^8$. The importance of correlation in the latter can be
seen from a study of Ce$^{4+}$, $4s^24p^64d^{10}5s^25p^6$~$^1S_0$ where occupied orbitals are excited by the SD process, to unfilled or unoccupied orbitals.
Variational calculations yielded a wave function expansion for which some of the larger basis states in $LSJ$ coupling are given in Table~\ref{tab:expand}. 
Of special interest are excitations {\it without a change in the principal quantum number} since they represent excitations between near-degenerate states of a complex. 

This investigation shows that the largest  excitation is $4d5p\rightarrow 4f5d$, namely a double excitation consisting of  single excitations from each of the two complexes. This is followed by $4d^2\rightarrow 4f^2$, and then $5p^2 \rightarrow 5d^2$ excitations.  The above contributions are too large to be considered as  a small correction for most applications. Also tested was the effect of adding the quadruple excitations $4d^2 LS \rightarrow 4f^2 LS$ and $5p^2 L'S' \rightarrow 5d^2 L'S'$ to the expansion. As shown in the Table, the coefficient for $LS=L'S'= \ ^3P$ was 0.0059, which might be important in some circumstances.
 
 Contributions to the wave function from
$4p^6$ or $4s^2$ are less than 0.0244 and 0.0173, respectively. Notice that all the large excitations within or between complexes did  not change their principal quantum number.

Ce$^{2+}$ differs in that the $n=4$ complex $4^{20}$ now has an extra unfilled subshell, $4f^2$, that leads to many states and the
analysis is not as simple but the concepts are the same. 

For Ce$^{4+}$, the (unnormalized) wavefunction generated from SD excitations of a super-complex can be written as
\begin{eqnarray}
\Psi(4^{18} 5^8\; ^1S_0) &=& \nonumber 
 \left[1  + \hat{S^2}(4) +  \hat{S^2}(5) + \hat{S^1}(4)\hat{S^1}(5)\right] \nonumber \\
& & \Phi( 4s^24p^64d^{10} \cdot 5s^25p^6 \; ^1\!S_0), 
\end{eqnarray} 
where $\hat{S^1}(n)$ and $\hat{S^2}(n)$  are the operators performing, respectively all single- and double- excitations among the designated $nl$ orbital set and, when applied to the configuration designating the complex, preserve parity and total quantum numbers. Here
we have used the fact that $\hat{S^1}(n)$ excitations by themselves are not allowed for $^1S_0$ states. The $\hat{S^1}(4)\hat{S^1}(5)$ excitation is a double excitation involving one orbital from each group.

A wave function of the form 
\begin{eqnarray}
\label{eqn:higher2} 
\Psi(4^{18} 5^8\; ^1S_0) &=& \nonumber 
[1  + \hat{S^2}(4)]\Phi(4s^24p^64d^{10}\; ^1\!S_0) \nonumber \\
&\times&  [1+ \hat{S^2}(5)] \Phi(5s^25p^6\; ^1\!S_0)  \\
&+&  \hat{S^1}(4)\hat{S^1}(5) \Phi( 4s^24p^64d^{10} \cdot 5s^25p^6 \; ^1\!S_0) 
\nonumber
\end{eqnarray} 
includes also some higher-order terms and would be appropriate when large
effects are present in both groups. Here the $\times$ operator represents the 
vector-coupling of CFS from the left set with those of the right and the required anti-symmetrization. Notice that in this form the correlation in the $n=4$ group is applied to each excitation of the $n=5$ group. If the size of the expansions are
$N_{4}, N_{5},$ and $N_{45}$ respectively, the number of basis states is $N_{4}\times N_{5} + N_{45}$.  When the expansion for $n=4$, for example, is fixed
then $N_{4} = 1$ and the expansion coefficients that need to be determined 
may reduce dramatically.

\section{ Ten valence electrons outside a $4d^{10}$ core}
\label{sec:valence}
In a {\sc Grasp} calculation, instead of complexes, the electrons are classified as inactive core, active core, and valence electrons.
In this study we treat $4d^{10}$ as an active core and $4f^25s^25p^6$ as 10 valence electrons. The $4s^24p^6$ subshells are relegated to the inactive core since the complex study showed their contribution to the energy was smaller. In these calculations SD excitations were applied to both $4f^25s^25p^6$ and $5s^25p^65d^2$ that define the MR set. Optimization was on all states of $4f^2$ weighted equally, with increasing orbital active sets up to $h$-orbitals but omitting $8h$. 
Orbital sets for $n=6-8$ were determined from interactions with the MR set as well as $4f5s^25p^65f$ in order to take into account any possible term dependence when the $4f$ orbitals were optimized separately.
The expansions for $n=8$ were extended to also include excitations from
the $4d^{10}$ core of each member of the MR set, expanded to include $4f^25s^25p^45d^2$, 
in order to 
estimate the effect of adding some CC correlation without any orbital optimization. The CC orbital set was limited to allow only excitations to $4f, 5d, 5f$ orbitals.
Results are shown in Table~\ref{tab:10-el}.

 \begin{table}[t]
\caption{$4f^2$ energy levels (in cm$^{-1}$) from a 10-electron calculation  compared with observed energy levels and the ground state energy ($E_g$ in Hartree units). The $n=8h$ results are extended to include an estimate of the core correlation in $+$CC. }
\label{tab:10-el}
\begin{center}
\begin{tabular}{r r r r r r r r}
\hline
  $LSJ$ &   $n=5$ &    $n=6h$ &      $n=7h$  &  $n=8h$  &   $+$CC      \\
        &        &           &             &         &   \\ 
\hline
  $^3H_4$ &     0 &       0 &         0 &         0 &         0   \\
  $^3H_5$ &  1400 &    1452 &      1464 &      1467 &      1619   \\
  $^3H_6$ &  2869 &    2905 &      2917 &      2919 &      3124   \\
  $^3F_2$ &  4375 &    4230 &      4123 &      4102 &      3859   \\
  $^3F_3$ &  5231 &    5092 &      4994 &      4976 &      4800   \\
  $^3F_4$ &  5357 &    5175 &      5047 &      5012 &      4752   \\
  $^1G_4$ &  7245 &    7045 &      6914 &      6878 &      6665   \\
  $^1D_2$ & 15130 &   14597 &     14232 &     14107 &     13522   \\
  $^3P_0$ & 19038 &   18475 &     18068 &     17931 &     17527   \\
  $^3P_1$ & 19376 &   18783 &     18386 &     18257 &     17708   \\
  $^3P_2$ & 20006 &   19377 &     18986 &     18859 &     18238   \\
  $^1I_6$ & 19531 &   19425 &     19308 &     19221 &     19634   \\
  $^1S_0$ & 40094 &   37501 &     36141 &     35386 &     33956   \\
  &&&&& \\
{$-E_g$} & 8848.58 & 8848.65 & 8848.66 & 8848.66 &   8848.82 \\
\hline
\end{tabular}
\end{center}
\end{table}

The results from these $n=5$ to $n=8h$ VV (valence) correlation calculations   have levels in their correct order. The fine structure for the $^3H$ has improved somewhat. Notice that the total energies have converged except for the highest level, namely $^1S_0$, for which convergence is slower. An investigation  of the wave function composition for $n=8h$ showed that the $4f^25s^25p^45d^2$ CSF had expansion coefficients larger than 0.09 for the $^1S_0$ state.  Comparison with the spectrum from the 2-electron study (Table \ref{tab:2-el}) shows that including correlation for the additional $5s^25p^6$ electrons has not had a large effect on the spectrum but did lower the total energy of the ground state by about 0.30~$E_h \approx 66,000~\mbox{cm} ^{-1}$. 
The largest effect is on the $^1S_0$ level.

\section{Contributions from the $4d^{10}$ core}

In the previous section,  $5s^25p^6$ was considered to be part of the valence electrons, with relatively small excitation energies. The  $4d^{10}$ electrons are different in that the $4d^2\rightarrow 4f^2$ excitation has a large effect on the total energy, although not on the $4f^2$ spectrum.

\subsection{Core-valence correlation}
In the super-complex of Ce$^{4+}$, a strong effect on the wave function composition  
arose  from the $4d5p \rightarrow 4f5d$ excitation.
In our computational method, such interactions are between core and valence electrons and account for the polarization of the $4d^{10}$ core by outer electrons.
Ce$^{2+}$ results are similar. The largest component arises from $4d^94f^35s^25p^55d$ and $4d^94f5s^25p^55d^3$ CSFs 
but many are small corrections that could be included as a first-order correction. 

\subsection{Properties of core correlation}

Core-correlation has some special properties in that all subshells are filled and have $^1S_0$ quantum numbers. Though {\sc Grasp} is fully relativistic, we will discuss this property in the non-relativistic case.

The SD excitations from the core shells of a  CSF consist of all excitations of the type $(ab)\pi LS \rightarrow (vv')\pi LS$  where $a,b$ are core orbitals, $\pi$ designates the parity of the pair of orbitals, and $vv'$ is any pair of unfilled or virtual orbitals. In the case of $4d^{10}$, the pairs can be derived by first uncoupling two equivalent electrons using the coupling relationship,
\small
 $$ |4d^{10}\; ^1S \rangle = \sum_{LS} |4d^8\; (LS).4d^2 (LS) \rangle  (d^{8}\; LS, d^2 \; LS |\}d^{10} \; ^1S) $$
\normalsize
 where 
$  (d^{8}\; LS, d^2 \; LS |\} d^{10} \; ^1S) $ is a coefficient of fractional grandparentage~\cite{Rac:43a}.
 The excited CSFs 
 are obtained by the replacement process $4d^2\; LS \rightarrow nln'l'\; LS$ . The possible $LS$ values for $d^2$ are $\{^1G,\; ^3F,\; ^1D,\;^3P,\; ^1S\}$ and these define the excited pair correlation functions for a correlated core.
 In the relativistic case, additional quantum numbers are needed as described in~\cite{Gaietal:2000a}.  The matrix element for the interaction from this excitation is the {\it same for all} CSFs, provided the $nln'l'$ orbitals are not present in the valence portion of the CSF. As a result, certain excitations may reduce the total energy (and affect the wave function) significantly but have a minor effect on a spectrum, since the latter is defined as an energy difference relative to the ground state. 
 
Core-correlation can be treated as a correction to an atomic state function
by correlating the core of all CSFs in the MR set.  This may be appropriate when the effect is small but for cases where the effect is large, the core of every CSF of the valence space should be correlated. 
One way of doing so is to use an effective Hamiltonian as is done in CI-MBPT~\cite{Kozetal:2015a}. In this case core-correlation is a first-order correction  of the wave function and is applied to {\it all} CSFs defining the valence space, including those that are introduced by the SD process. At no point are the interactions between these corrections introduced. A more general approach is given by Eq.~\ref{eqn:higher2}.

\subsection{Results for an active $4d^{10}$ core} 

\begin{table}[t]
\caption{$4f^25s^25p^6$ energy levels (in cm$^{-1}$)  from calculations that include correlation with the $4d^{10}$ active core. Also reported is the total energy $-E_g$ of the ground state and the number (No.) of CSFs in the expansions.}
\label{tab:20-el}
\begin{center}
\begin{tabular}{r r r r r r r r}
\hline
  $LSJ$ &   $n=4$ &  $n=5$ & $n=6$ & $n=7$ & ASD  &   CI all \\
             &         &          &       &     &  \cite{NIST_ASD} & \cite{Safetal:2015a} \\
\hline
    $^3H_4$ &         0 &         0 &         0 &         0 &         0 &  0 \\
    $^3H_5$ &      1636 &      1516 &      1598 &      1593 &   1528.32 & 1565 \\
    $^3H_6$ &      3296 &      3116 &      3233 &      3204 &   3127.10 & 3227 \\
    $^3F_2$ &      4685 &      4250 &      4305 &      4299 &   3762.75  \\
    $^3F_3$ &      5749 &      5321 &      5393 &      5371 &   4764.76  \\
    $^3F_4$ &      7899 &      5496 &      5513 &      5477 &   5006.06  \\
    $^1G_4$ &      5680 &      7542 &      7620 &      7555 &   7120.00 & 7650 \\
    $^1D_2$ &     16693 &     15350 &     15242 &     15109 &  12835.09 & 13786 \\
    $^3P_0$ &     21043 &     19059 &     19053 &     18941 &  16072.04  \\
    $^3P_1$ &     21541 &     19451 &     19408 &     19264 &  16523.66  \\
    $^3P_2$ &     22411 &     20138 &     20118 &     19953 &  17317.49  \\
    $^1I_6$ &     23391 &     20158 &     19992 &     19829 &  17420.60  \\
    $^1S_0$ &     41547 &     40504 &     39452 &     38758 &  32838.62  \\
    &&&&& \\
\multicolumn{1}{l}{$-E_g$  } &  8848.62 &   8848.85 &   8849.03 &   8849.07\\
\multicolumn{1}{l}{ No.} &  33\,520 & 1\,606\,947 & 2\,678\,670 & 4\,679\,330\\
\hline
\end{tabular}
\end{center}
\end{table} 

Table~\ref{tab:20-el} shows some results for calculations that include VV, CV, and CC correlation effects on the $4f^2$ spectrum with an active $4d^{10}$  core.  Expansions increase in size rapidly so the orbital set for CC needs to be controlled as well as the MR set. In the $n=4$ calculation, the MR set included both $4f^2 5s^2 5p^6$ and $5s^2 5p^6 5d^2$ and an orbital set with orbitals up to $\{5s,5p,5d,4f\}$ or simply $\{5554\}$. The inactive core orbitals were the same as those of the 2-electron calculation. Excitations were SD excitations from all shells. Double excitations from $4d^{10}$ were limited to
excitations to $\{4f, 5d, 5f\}$ orbitals with the $5g$ orbital participating only in CV and VV in the $n=5$ calculation with a \{55555\} excitation orbital set. The MR set now also contained $4f5s^25p^65f$, $4f 5s^2 5p^5 5d^2$, and $5s^25p^45d^4$,
although the latter two did not contribute to CC, the number of excitations being too numerous for inclusion.
The effect of including CC was the contraction of the $(4f_-,4f_+)$ orbitals from a mean radius of (1.174,1.189) $a_0$ to (1.095, 1.091) $a_0$.
The fine-structure splitting of the lowest term is now in excellent agreement with observation. The $n=5$ expansion was reduced by extracting those CSFs with an expansion coefficient greater in magnitude than 0.00001 in at least one eigenvector. To this were added CSFs from an $n=6$  expansion including at least one $n=6$ orbital in a CV+VV expansion from the five members of the MR set. Again, the $n=6$ results were reduced and $n=7$ CSFs added to the reduced expansion.  The new CSFs have had a
small effect on the lower levels but make a significant contribution to higher levels. Note that the $^3P$ fine-structure is in fairly good agreement with observation in that all levels of the latter are shifted by a similar amount.  At the same time, comparing the final ground state energy for the 10-electron system reported in Table~\ref{tab:10-el}, the ground state energy has been lowered by 0.25~$E_h$ or $61,453~\mbox{cm}^{-1}$. In other words, correlation shifts the total energies more than it modifies the spectrum.

Except for the $^1D_2$ level, the lower levels of the $n=5$ calculation agree with observation slightly better than the best results reported by Safronova {\it et al}~\cite{Safetal:2015a}.

\section{Analysis}

\begin{table}[t]
\caption{Analysis of the wavefunction composition and total energy (in $E_h$)  for the $4f^2$ $^1S_0$ state from  the three types of calculations. Included is the expansion coefficient (Coef) and the CSF when converted to $LSJ$ coupling. }
\label{tab:wf-comparison}
\begin{center}
\begin{tabular}{r l}
\hline 
Coef.  &   CSF  \\
\hline 
\multicolumn{2}{l}{$4f^2$ $^1S_0$:  $E= -8848.2204$ } \\
         0.8522 & $4f^2$ $^1S_0$ \\
         0.4794 & $5d^2$ $^1S_0$ \\
        $-$0.1351 & $4f5f$ $^1S_0$ \\
        $-$0.1246 & $4f^2$ $^3P_0$ \\
        $-$0.0638 & $5f^2$ $^1S_0$ \\
&  \\
\multicolumn{2}{l}{$4f^25s^25p^6$ $^1S_0$:  $E= -8848.5021$} \\
         0.9020 &   $4f^2(^1S)5s^25s^25p^6$ $^1S_0$  \\
         0.2780 &   $5s^25p^65d^2$ $^1S_0$   \\
        $-$0.0903 &   $4f^2(^1S)5s^25p^4(^3P)5d^2(^3P)$ $^1S_0$ \\
        $-$0.0864 &   $4f^2(^3P)5s^25p^6$ $^3P_0$ \\
        $-$0.0820 &   $4f^2(^1S)5s^25p^4(^1D)5d^2(^1D)$ $^1S_0$ \\
        $-$0.0798 &   $4f5s^25p^66f$ $^1S_0$ \\
        $-$0.0719 &   $ 5s^25p^65d6d$ $^1S_0$ \\
        $-$0.0696 &   $ 4f5s^25p^65f$ $^1S_0$ \\
         0.0695 &   $ 4f^2(^1S)5s^25p^4(^1S)5d^2(^1S)$  $^1S_0$ \\
        $-$0.0652 &   $ 4f5s^25p^5[^1D]5d^2(^1D)$ $^1S_0$ \\
        $-$0.0645 &   $ 4f5s^25p^5[^1G]5d^2(^1G)$ $^1S_0$ \\
         0.0594 &   $ 4f^2(^1D)5s[^2D]5p^65d$ $^1S_0$ \\
        $-$0.0531 &   $ 4f^3(^2F)5s[^3F]5p^5[^2D]5d$ $^1S_0$ \\
        $-$0.0455 &   $ 4f^3(^2P)5s^25p^5$ $^1S_0$\\
        $-$0.0448 &   $ 5s^25p^65f^2(^1S)$ $^1S_0$ \\
&  \\
\multicolumn{2}{l}{$4d^{10}4f^25s^25p^6$ $^1S_0$:  $E= -8848.8937$} \\
         0.9029 &  $4d^{10}4f^2(^1S)5s^25s^25p^6$ $^1S_0$  \\
         0.2579 &  $4d^{10}5s^25p^65d^2$ $^1S_0$   \\
        $-$0.1263 &  $4d^{10}4f5s^25p^65f$ $^1S_0$ \\
        $-$0.0865 &  $4d^{10}4f^2(^3P)5s^25p^6$ $^3P_0$ \\
        $-$0.0721 &  $ 4d^{10}4f^2(^1S)5s^25p^4(^3P)5d^2(^3P)$ $^1S_0$ \\
        $-$0.0667 &  $ 4d^{10}4f^2(^1S)5s^25p^4(^1D)5d^2(^1D)$ $^1S_0$ \\
        $-$0.0623 &  $ 4d^{10}4f5s^25p^5[^1G]5d^2(^1G) $ $^1S_0$ \\
         0.0610 &  $ 4d^94f^3(^2F)[^1P]5s^25p^5[^2D]5d$ $^1S_0$ \\
        $-$0.0587 &  $ 4d^{10}4f5s^25p^5[^1D]5d^2(^1D)$  $^1S_0$ \\
         0.0558 &  $ 4d^{10}4f^2(^1S)5s^25p^4(^1S)5d^2(^1S)$ $^1S_0$ \\
         0.0556 &  $ 4d^{10}4f5s^25p^66f$ $^1S_0$ \\
        $-$0.0478 &  $ 4d^8(^3P)4f^4(^3P)[^1S]5s^25p^6$  $^1S_0$ \\
         0.0474 &  $ 4d^{10}4f^2(^1D)5s[^2D]5p^65d$  $^1S_0$ \\
         0.0462 &  $ 4d^8(^1S)4f^4(^1S)5s^25p^6$ $^1S_0$ \\
\hline
\end{tabular}
\end{center}
\end{table}

Comparison of computed energy levels with those derived from observation is a common method for assessing the accuracy of a calculation. But, as we have already seen, not all contributions to a wave function affect the computed spectrum.  
 For the prediction of other atomic properties such as lifetimes or transition rates, the accuracy of the wave function composition is a more important factor. For the analysis of a wave function it is convenient to transform the expansion to $LSJ$ coupling~\cite{Gaietal:2017a}. The expansion coefficients depend on the radial basis but a wavefunction can also be viewed as a linear combination of multi-electron spin-angular functions that are not affected by radial transformations.
 
Table~\ref{tab:wf-comparison} shows how the expansion coefficients for major
contributors to the $4f^2\;^1S_0$ wave function change with the correlation model. Given are the coefficients of some CSFs (the contribution to the composition is the square of the  coefficient) for the three approximations -- the $4f^2$ two-electron system outside inactive closed shells, the $4f^25s^25p^6$ 10-electron system outside closed shells, and finally the $4d^{10}4f^25s^25p^6$ 20-electron system outside  closed shells. For the first method, there is strong interaction between $4f^2$ and $5d^2$ partly because the $5d^2$ energy levels 
overlap those of $4f^2$ and the energy difference of the two is too small. The lowest $5d^2$ level is $^3F_2$ (not included in any Table) and its computed energy level is 33~558 cm$^{-1}$ compared with the observed value of 40~440.20 cm$^{-1}$.
Including the correlation of $4f^2$ with $5s^25p^6$, increases the separation between the levels and reduces the expansion coefficient. Including also the correlation with $4d^{10}$ further decreases the contribution to the wave function  by a relatively small amount. At the same time, the computed $4d^2~^3F_2$ energy level is now 63~429~cm$^{-1}$ 
and hence too high. 

Table~\ref{tab:wf-comparison} also shows that the core correlation lowers the total energy of the $^1S_0$ level by slightly more than correlation between $4f^2$ and the $5s^25p^6$ closed shells in that the difference in total energies of the state is slightly larger between the last two results than the first two. Because the number of SD excitations from $4d^{10}$ increases extremely rapidly with the size of the excitation orbital set, the present work has limited its size. 
As in the super-complex discussed earlier, the largest excitation is $4d5p\rightarrow 4f5d$ but with a smaller expansion coefficient, namely 0.0610 compared with 0.0854 for the complex, as shown in Table~\ref{tab:expand}. Similarly, other excitations also have smaller coefficients which may be related to the presence of the $4f^2$ electrons but may also be the result of correlating the core of only a few CSFs which has the effect of increasing energy differences and thereby decreasing the expansion coefficients. Further studies are needed.

\section{Conclusion}

Accurate predictions for Lanthanide spectra with multiple open shells provide a challenge for theory. In this work, results were based on the  {\sc Grasp} code that computes a wave function from an MR set along with SD excitations from members of this set, thus including  selected higher-order terms and resulting in expansions with millions of basis states. 

In effect, correlation is a local phenomenon arising from corrections to the wave from the $\sum 1/r_{ij}$ singularities in the Hamiltonian, but orbitals are global in nature making the calculations difficult, mainly because of the number of basis states. In Ce$^{2+}$, ignoring the inactive subshells, there are three correlation regions for which correlation can be computed without difficulty in GRASP, namely $\Psi(4s^2 4p^6 4d^{10} \; ^1S_0)$, $\Psi(4f^2 \; J\pi)$, and $\Psi(5s^2 5p^6 \; ^1S_0)$ where each $\Psi$ is an expansion over CSFs. Then, following the concepts first introduced by Chung~\cite{Chu:91a} and applied successfully to Be-like systems~\cite{Chuetal:93a}, the wave function for Ce$^{2+}$  becomes

\begin{eqnarray}
 && \hspace*{-1cm} \Psi(4s^2 4p^6 4d^{10} 4f^2 5s^2 5p^6 \;  \pi J) \nonumber \\
& = & \Psi(4s^2 4p^6 4d^{10} \;  ^1S_0)\times \Psi(4f^2 \;  \pi J)\times \Psi(5s^2 5p^6 \; ^1S_0) \nonumber \\
 & + & \  \hat{S}_{2o3}  \; \Phi( 4s^2 4p^6 4d^{10} \cdot 4f^2 \cdot 5s^2 5p^6 \;  \pi J)  
\end{eqnarray}

\noindent
where the three individual expansions $\Psi$ are vector coupled and anti-symmetrized similar to the way in which  CSFs for a group of subshells are vector coupled. 
The last term represents the CSF expansion produced by an excitation operator $\hat{S}_{2o3}$ involving at least two of the three ($2o3$) subgroups separated by a centered dot.
Excluded  are excitations for which all excitations are from the same subgroup. This equation is directly related to the equation for generating expansions for a super-complex, namely Eq.~\ref{eqn:higher2}, but here the limitation on excitations has been removed and the equation is not restricted to SD.  The fastest rate of convergence for each group would require a different orbital basis for each leading to a non-orthogonal basis for the full wave function. The present version of GRASP assumes one orthonormal orbital basis leading to larger expansions whose size would be the product of the three sizes. But this partitioned approach could also provide valuable information about when higher-order excitations such as TQ excitations are  needed.

 In the present case, the configuration $4d^84f^45s^25p^45d^2$ is the coupled product of excitations $4d^2\rightarrow 4f^2$ and $5p^2\rightarrow 5d^2$, a special case of a quadruple excitation. From Table~\ref{tab:wf-comparison} we see that the largest expansion coefficient in $^1S_0$ is $-$0.0721 for the $5p^2 \rightarrow 5d^2$ excitation whereas the largest coefficient is $-$0.0478 for the $4d^2\rightarrow 4f^2$ expansion. 
Depending on the accuracy required for the wavefunction, the higher-order term may be needed.  At the same time, as shown earlier, matrix elements for core correlation may be the same for many CSFs. For example, the CC excitation $4d^2\rightarrow 4f^2$ of a given $\pi J$ produces a matrix element for the interaction that is the same for all CSFs that do not already include a $4f$ orbital in their definition.  The present code treats each matrix element independently.  

A reorganization of the way core-correlation is included in GRASP has the possibility of greatly improving the efficiency of the program for lanthanides and other heavy elements.
 
\begin{acknowledgments}
\noindent 
The authors (CFF and MRG) acknowledge support from the Canada NSERC Discovery Grant 2017-03851 and 
 the FWO \& FNRS Excellence of Science Programme (EOS-O022818F), respectively.
\end{acknowledgments}


%

\end{document}